\documentclass[prb,twocolumn,aps,showpacs,fixfloats]{revtex4-1}
\usepackage{graphicx}
\usepackage{bm}
\usepackage{amsmath,amssymb}
\usepackage{subfigure}
\usepackage{float}
\usepackage{latexsym}
\usepackage{epstopdf} 
\usepackage{color}
\usepackage{pdfpages}

\newcommand{\s}{\scriptscriptstyle}

\begin{document}

\title {Spin pumping from a ferromagnet into a hopping insulator: the role of resonant absorption of magnons  }

\author{Z. Yue, D. A. Pesin,  and M. E. Raikh }

\affiliation{Department of Physics and
Astronomy, University of Utah, Salt Lake City, UT 84112, USA}

\begin{abstract}
Motivated by recent experiments on spin pumping from a ferromagnet into organic materials
in which the charge transport is due to hopping, we study theoretically the generation and propagation of spin current in a hopping insulator. Unlike metals, the spin polarization
at the boundary with ferromagnet is created as a result of magnon absorption within  pairs of localized states and it spreads following the current-currying  resistor network (although the charge current is absent). We consider a classic resonant  mechanism
of the ac absorption in insulators and adapt it to the absorption of magnons.
A strong enhancement of pumping efficiency is predicted when the Zeeman splitting of the localized states in external magnetic field is equal to the frequency of ferromagnetic resonance. Under this condition the absorption of a magnon takes place within {\em individual} sites.

\end{abstract}

\maketitle

%\section{Introduction}
\begin{figure}
\includegraphics[width=70mm]{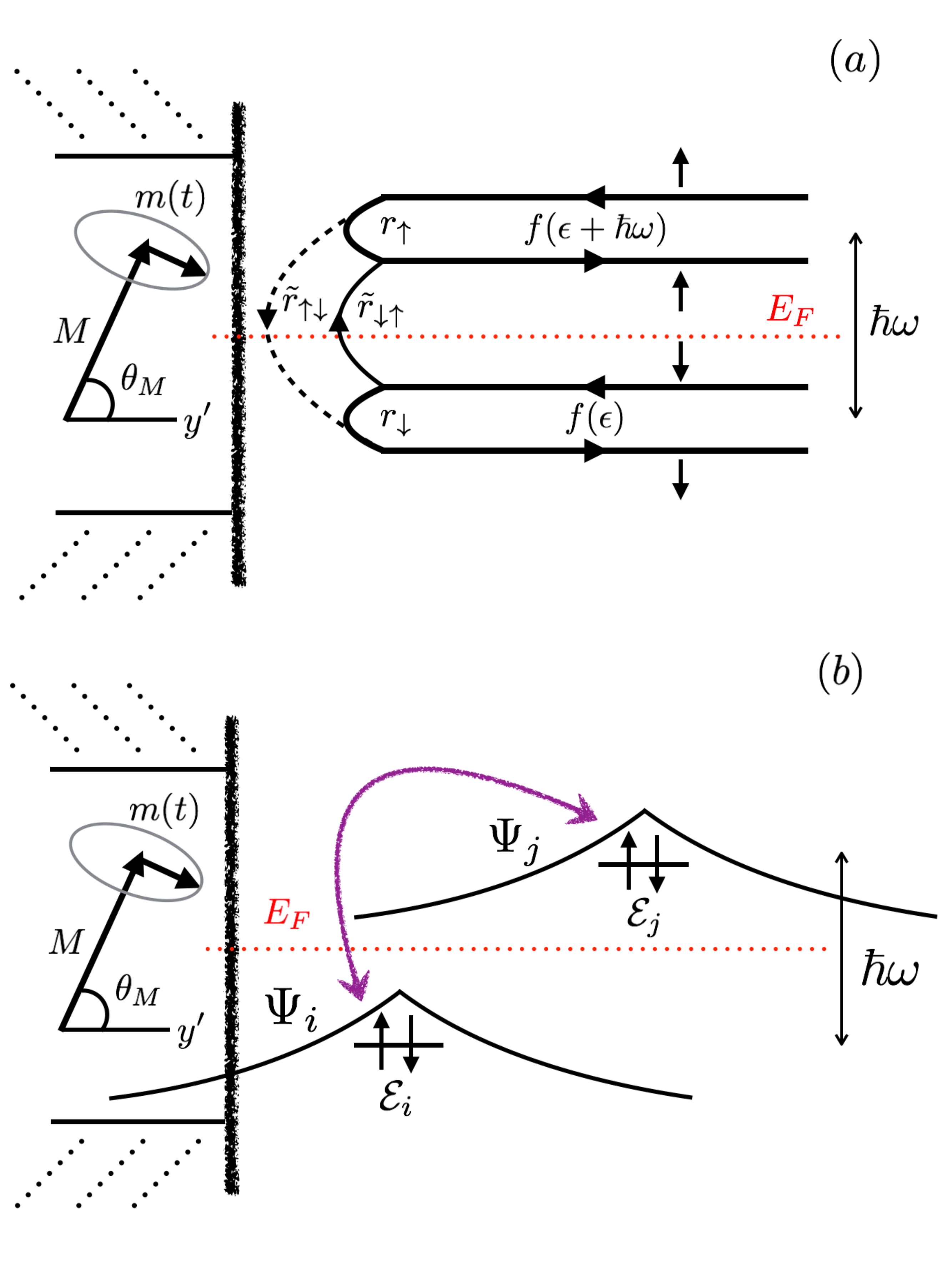}
\caption{(Color online) Elementary processes underlying the spin pumping into a metal (a), and into an insulator (b).  In the metal, an $\uparrow$   electron, impinging on the N-F boundary, is primarily reflected elastically with amplitude $r_{\uparrow}$. Spin precession in F gives rise to inelastic reflection  with amplitude ${\tilde r}_{\uparrow\downarrow}$ associated with the emission of a magnon. A $\downarrow$ electron is either reflected elastically with amplitude $r_{\downarrow}$,
or inelastically, after absorbing of a magnon, with amplitude ${\tilde r}_{\downarrow\uparrow}$.
 The injected spin current is proportional to $|{\tilde r}_{\uparrow\downarrow}|^2\omega \frac{\partial f}{\partial \varepsilon}$. In the insulator,
only inelastic processes are at work. Emission and absorption of magnons take place within pairs
of localized states.}
\label{Fig1}
\end{figure}

\section{Introduction}

The phenomenon of spin pumping from a ferromagnet (F) into a normal (N) layer is one of the most prominent approaches to the generation of pure spin currents.
A prime manifestation that  pumping indeed takes place in realistic F-N structures is the additional broadening\cite{Tserkovnyak2002} of the ferromagnetic resonance (FMR)
in F, caused by a contact with N-layer. This additional broadening was first observed experimentally in Ref.~[\onlinecite{FromDamping}]. Another, more delicate, manifestation of pumping was reported shortly after.
Namely, the injected spin current, entering the nonmagnetic material with spin-orbit coupling (like Pt) causes a voltage drop across the current direction. This voltage drop is due to
the inverse spin-Hall effect\cite{book1} (ISHE), and has a maximum when the frequency of the microwave radiation driving the ferromagnet, $\omega$, is equal to the FMR frequency, $\omega_{\s \textit{FMR}}$. Pioneering observations of  pumping via ISHE in Refs.~[\onlinecite{Pt1,Pt2,Pt3}] utilized Pt as the normal layer.\cite{Pt4,Pt5,Pt6} They were followed by
reports on similar observations of pumping into different materials\cite{Bader,Gold,ZnO,Exotic}, including prominent semiconductors GaAs\cite{GaAs}, Si\cite{silicon1,silicon2}, Ge\cite{germanium}, and, most recently,
graphene.\cite{graphene} Experimental results on the electric field generated due to ISHE , ${\bm E}_{\s \textit{ISHE}}$,
are analyzed  using the relation ${\bm E}_{\s \textit{ISHE}}\propto {\bm J}^{(s)}\times {\bm \sigma}$, where ${\bm J}^{(s)}$ determines the spatial direction of the spin current flow and its magnitude, while ${\bm \sigma}$ is its polarization. The magnitude of the spin current is given by
\begin{equation}
\label{relation}
J^{(s)}= g_{\uparrow\downarrow}C\Big[{\bm m}(t)\times\frac{d {\bm m}(t)}{d t}\Big]_z,
\end{equation}
where $z$ axis is taken along the static part of the magnetization. In Eq. (\ref{relation}) the constant $C$ characterizes the properties of the normal layer (like ratio of thickness to the spin-diffusion length), while ${\bm m}(t)$ describes the magnetization dynamics in the  ferromagnet. The expression for  ${J}^{(s)}$ has the same form as the damping term in the equation that governs ${\bm m}(t)$. It was a
remarkable experimental finding\cite{Pt2}  that ISHE voltage exhibits essentially the same behavior as a function of microwave power and the deviation of $\omega$  from
$\omega_{\s \textit{FMR}}$ as the additional FMR damping.

Microscopic physics of pumping is encoded in the mixing constant\cite{Tserkovnyak2002,Nazarov,Review} $g_{\uparrow\downarrow}$ in Eq. (\ref{relation}). A fundamental process underlying the pumping is the inelastic electron-magnon scattering at the F-N interface. Microscopic treatment of this scattering\cite{Zangwill2002,Microscopic2014}
assumes that electrons of the normal layer impinging on the interface with ferromagnet are {\em plane waves}. On the other hand, in a number of recent papers\cite{Polaron1,Polaron2,Polaron3,Polaron4}  spin pumping into organic materials sandwiched between ferromagnet and Pt has been reported. Strong temperature dependence of the resistance in these materials\cite{ActivationEnergy}
suggests that the charge transport is due to hopping of polarons\cite{Polaron2,Polaron4}, so that the description of pumping based on plane waves does not apply. This raises the question about the microscopics of spin pumping  in the localized regime.

In the present paper we consider theoretically the spin pumping into a hopping insulator using the minimal model of coupling of localized states to a ferromagnet. We demonstrate that, unlike metals,
the underlying process responsible for pumping is the resonant magnon absorption accompanied by transitions between localized states, see Fig. \ref{Fig1}. A distinctive feature of pumping into an insulator is that that the pumping efficiency, commonly described by a constant, $g_{\uparrow\downarrow}$,  depends strongly on the external dc magnetic field. This is because, in addition to causing the spin precession in  ferromagnet, this field modifies the spin structure of the localized states between which the magnon is absorbed, see Fig. \ref{Fig1}.  The effect of external field is most pronounced when the waiting time for a hop
is longer than the period of the ac  field which drives the FMR.
Since the resonance frequency, $\omega_{\s \textit{FMR}}$, depends on the orientation of the external field\cite{Kittel}, for certain
orientations\cite{Silsbee1979} this frequency coincides with the Zeeman splitting of the localized states, Fig. \ref{Fig2}. Spin pumping is most efficient for such orientations, since the absorption of magnon takes place within  {\em individual} sites.
We also show that, with no charge current, the spin polarization generated at the F-N boundary, spreads in the insulator along the same percolation network\cite{HalperinLanger,book} that determines the electrical resistance.

\section{Absorption of magnons at F-N boundary}

\subsection{General considerations}

\begin{figure}
\includegraphics[width=70mm]{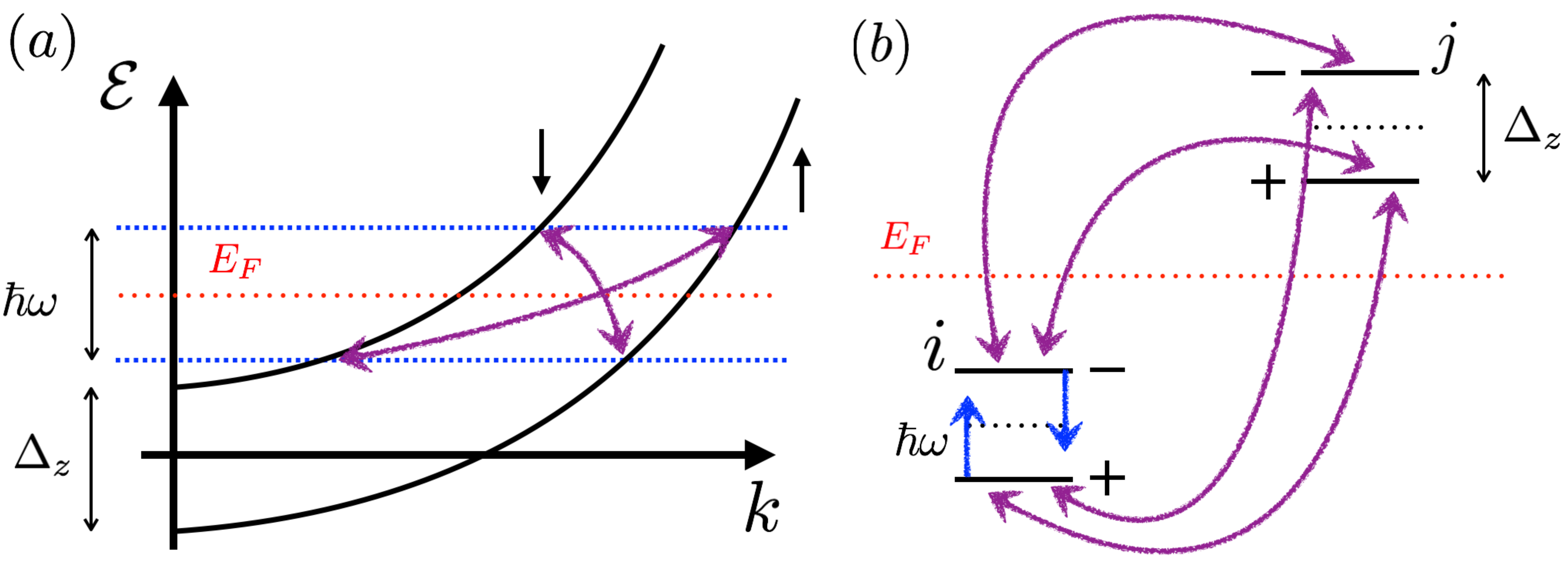}
\caption{(Color online) Illustration of pumping in metal (a) and in insulator (b) in the
presence of a Zeeman splitting, $\Delta_z$. In metal, the absorption  (emission) of a magnon, $\hbar \omega$, near the F-N boundary does not conserve momentum, and thus is insensitive to the
ratio $\Delta_z/\hbar\omega$. By contrast, in and  insulator, and near the condition
$\hbar \omega=\Delta_{z}$, the absorption (emission) of a magnon is {\em resonant}.
}
\label{Fig2}
\end{figure}

Figure~\ref{Fig2} illustrates the difference between pumping into a metal, and into an insulator
in an applied magnetic field, ${\bm H}$. While  ${\bm H}$ is responsible for the magnetization precession precession in the ferromagnet, it also causes a spin splitting, $\Delta_{z}$, of the spectrum
in the metallic normal layer, Fig. 2a. This splitting, however, does not affect the absorption of magnons. The reason is that the absorption at a boundary does not require momentum conservation, i.e. the matrix element is constant, and thus there is no dependence of the spin current, $I^{(s)}$, on the dc field in the
normal layer.
%${\bm H}$.

The situation is different for an insulator, where the magnon absorption takes place between the discrete levels, Fig. \ref{Fig2}(b). In this case, and for a general orientation of ${\bm H}$, the Zeeman levels are the linear combinations of $\uparrow$ and $\downarrow$ spin states. As a result, transitions from {\em each} of the initial states on site $i$ to {\em both} final states on site $j$ are allowed. This fact distinguishes absorption of magnons from the conventional absorption
of an ac {\em electric} field \cite{ResonantAbsorption,Efros,WithCoulomb}, and, as we will see below, gives rise to ${\bm H}$-dependence of the spin current.
Another origin of ${\bm H}$-dependence is the possibility of {\em intrasite} absorption
of magnons at the boundary. We will see that the intrasite transitions dominate the absorption
near the resonant condition $\hbar\omega=\Delta_{z}$. Away from this condition, the intersite transitions dominate.

\subsection{The model}

Consider a pair of localized states, $i$ and $j$, Fig. \ref{Fig2}(b).
%, belonging to the current-carrying
%network.
Assume for simplicity that the ferromagnet is an insulator, i.e. it is a barrier for electrons in N. Precession, ${\bm m}(t)$,  of magnetization in ferromagnet  can be modeled as a time-dependent correction $\propto {\bm m}(t){\hat \sigma}$ to the barrier potential.
The pumping takes place since the wave function, $\Psi_i$, can penetrate under the barrier. As
a result, the Hamiltonian of site $i$ has a correction
\begin{equation}
\label{deltaHi}
\delta{\hat H}_i=J \Bigl[{\hat \sigma}_x m_x\sin\omega t+ {\hat \sigma}_y m_y \cos\omega t\Bigr],
\end{equation}
where $J$ accounts for tunneling. Projections $m_x(t)$ and $m_y(t)$ are proportional to the
magnitude of the microwave field and depend in a resonant way on the proximity of $\omega$ to
$\omega_{\s \textit{FMR}}$. Analytical expressions for these projections can be found e.g. in Ref.~[\onlinecite{Pt6}].

The Hamiltonian $\delta{\hat H}_i$ of Eq.~(\ref{deltaHi}) causes transitions of electrons between the sites $i$ and $j$.
Absorption of energy in course of these transitions is quite similar to the absorption of the ac {\em electric} field by pairs of the localized states. However, the transitions caused by $\delta{\hat H}_i$ are accompanied by  spin flips, both from $\uparrow$ to $\downarrow$, and from $\downarrow$ to $\uparrow$. With regard to absorption of energy, one should add up the contributions of the both types of transitions, i.e.
\begin{equation}
I^{(e)}=I_{\downarrow \rightarrow \uparrow}+I_{\uparrow \rightarrow \downarrow}
\end{equation}
However, the spin current results from the fact that these contributions are not equal to each other, so that

\begin{equation}
\label{SpinCurrent}
I^{(s)}=I_{\downarrow \rightarrow \uparrow}-I_{\uparrow \rightarrow \downarrow}.
\end{equation}
Thus, for calculation of the spin current into hopping insulator, one
can use the standard ``resonant" phononless absorption theory\cite{ResonantAbsorption}
and substitute the corresponding rates into Eq. (\ref{SpinCurrent}).

\subsection{Resonant absorption at ${\bm H}=0$}
%\noindent{\em Resonant absorption.}
We first neglect the Zeeman splitting in the normal layer. In this case resonant transitions happen within pairs of localized states, Fig.~\ref{Fig1}b. The correction $\delta{\hat H}_i$ causes such transitions between the sites $i$ and $j$ because the corresponding wave functions $|i\rangle$ and $|j\rangle$ have a finite overlap integral, ${t}_{ij}$.\cite{Efros} Due to this overlap, the eigenfunctions of the pair get modified as
\begin{eqnarray}
  |{\bf i} \rangle&=&\sqrt{\frac{\Gamma+\delta\varepsilon}{2\Gamma}}|i\rangle+\sqrt{\frac{\Gamma-\delta\varepsilon}{2\Gamma}}|j\rangle,\nonumber\\
  |{\bf j}\rangle&=& -\sqrt{\frac{\Gamma-\delta\varepsilon}{2\Gamma}}|i\rangle+\sqrt{\frac{\Gamma+\delta\varepsilon}{2\Gamma}}|j\rangle,
\end{eqnarray}
for $\delta\varepsilon=\varepsilon_j-\varepsilon_i>0$. The corresponding energies are
 %With these definitions, $|s\rangle\sim|i\rangle$ and $|a\rangle\sim|j\rangle$ for $\varepsilon_j-\varepsilon_i\gg|{t}_{ij}|>0$, while $|s\rangle\sim|j\rangle$ and $|a\rangle\sim|i\rangle$ for $\varepsilon_i-\varepsilon_j\gg|{t}_{ij}|>0$.

%\begin{equation}
%|\text{\bf i}\rangle=\frac{{\tilde\varepsilon_i}-{\tilde\varepsilon_j}}{\Gamma}|i\rangle+\frac{2{t}_{ij}}{\Gamma}|j\rangle
%,~~~~|\text{\bf j}\rangle=\frac{2{t}_{ij}}{\Gamma}|i\rangle+\frac{{\tilde\varepsilon_j}-{\tilde\varepsilon_i}}{\Gamma}|j\rangle,
%\end{equation}
\begin{equation}
{\tilde\varepsilon}_{i,j}=\frac{\varepsilon_i+\varepsilon_j}{2}\mp\frac{\Gamma}{2}
,~~~~\Gamma=\Bigl[\delta\varepsilon^2+4{t}_{ij}^2\Bigr]^{1/2}.
\end{equation}
Since both modified eigenfunctions contain $|i\rangle$, the matrix element of $\delta{\hat H}_i$ between them is finite, and the Golden-rule expression for the spin-flip part of the $i\rightarrow j$ transition rate for $\varepsilon_j>\varepsilon_i$ reads
\begin{equation}
\label{current}
I^{(s)}_{i\rightarrow j}=-m_x m_y J^2{\bf\mathrm{F}}({\tilde\varepsilon_i},{\tilde\varepsilon_j},\omega),
\end{equation}
where the function ${\bf\mathrm{F}}$ is defined as
\begin{align}
\label{functionF}
 {\bf\mathrm{F}}({\tilde\varepsilon_i},{\tilde\varepsilon_j},\omega)&=\frac{2{t}_{ij}^2}
 {({\tilde\varepsilon}_j-{\tilde\varepsilon}_i)^2}
\frac{\frac{1}{\tau}[f({\tilde\varepsilon_i})-f({\tilde\varepsilon_j})]}
{({\tilde\varepsilon}_j-{\tilde\varepsilon}_i-\hbar\omega)^2+\left(\frac{\hbar}{\tau}\right)^2} \nonumber
 \\ &=\frac{2{t}_{ij}^2}{\Gamma^2}
\frac{\frac{1}{\tau}[f({\tilde\varepsilon_i})-f({\tilde\varepsilon_j})]}
{(\Gamma-\hbar\omega)^2+\left(\frac{\hbar}{\tau}\right)^2}.
\end{align}
%\Bigl[\frac{({\tilde\varepsilon_i}-{\tilde\varepsilon_j}) I_{ij}}{\Gamma^2}\Bigr]^2
%\frac{[f({\tilde\varepsilon_i})-f({\tilde\varepsilon_j})]\frac{1}{\tau}}
%{(\Gamma-\omega)^2+\frac{1}{\tau^2}}.
Here we have introduced the phonon broadening of the levels, $\tau^{-1}$.

It is easy to see that the transition rate to states with $\varepsilon_j<\varepsilon_i$ is given by Eq.~(\ref{current}) with function ${\bf\mathrm{F}}$ from Eq.~(\ref{functionF}), but with $f({\tilde\varepsilon_i})\leftrightarrow f({\tilde\varepsilon_j})$, and thus the rate has the same sign as Eq.~(\ref{current}). Physically, this can be seen from the following argument: Consider the simple case of $m_x=m_y$. The Hamiltonian of Eq.~(\ref{deltaHi}) implies that for a given site at the interface, spins $\uparrow$ are transferred to states of higher energy (and there is a backflow of spins $\uparrow$ converted from $\downarrow$ from those states), while spins $\downarrow$ are pushed to states with lower energy (and there is a backflow of spins $\downarrow$ converted from $\uparrow$). Since the occupation of the state at the interface is larger than of those at higher energy, there is a negative $\downarrow\to \uparrow$  conversion rate because of transitions up the energy. This is exactly what Eq.~(\ref{current}) suggests. Further, since the occupation of the state at the interface is lower than of those at lower energy, there is a positive $\uparrow\to \downarrow$ conversion rate, or, again, negative $\downarrow\to \uparrow$  one. Hence a simple permutation $f({\tilde\varepsilon_i})\leftrightarrow f({\tilde\varepsilon_j})$ suffices to describe transitions to states with $\varepsilon_j<\varepsilon_i$.

The product $m_xm_y$ in Eq. (\ref{current})
is specific for spin pumping, see Eq. (\ref{relation}). The expression for the net absorption rate contains $\frac{1}{2}(m_x^2+m_y^2)$ instead. Another difference from the conventional resonance absorption\cite{book,ResonantAbsorption} is
the structure of the matrix element in Eq. (\ref{current}).  This, however, modifies the result of averaging over the sites, $j$,
only by a numerical factor. A crucial observation in the averaging procedure\cite{ResonantAbsorption} is that the relevant sites, $j$, are located within a narrow
spherical layer with a radius $r_{\omega}$ which is found from the condition $2|{t}_{ij}(r_{\omega})|=\hbar\omega$. Assuming the exponential decay of the overlap integral with distance, $|{t}_{ij}(r)|={t}_0\exp(-r_{ij}/a)$, we have
\begin{equation}
  r_{\omega}=a\ln\frac{2{t}_0}{\hbar\omega}.
\end{equation}
The result of averaging and summing over sites far away from the boundary reads
\begin{equation}
\label{ZeroField}
I^{(s)}(\omega)=2\pi^2 m_x m_yJ^2\left(g\omega ar_{\omega}^2\right)\frac{\partial f}{\partial \varepsilon},
\end{equation}
where $g$ is the density of states. The transition rate of Eq.~(\ref{ZeroField}) should be interpreted as the spin current generated per a localized state coupled to the ferromagnet.

\subsection{Resonant absorption at finite ${\bm H}$}

\begin{figure}
\includegraphics[width=70mm]{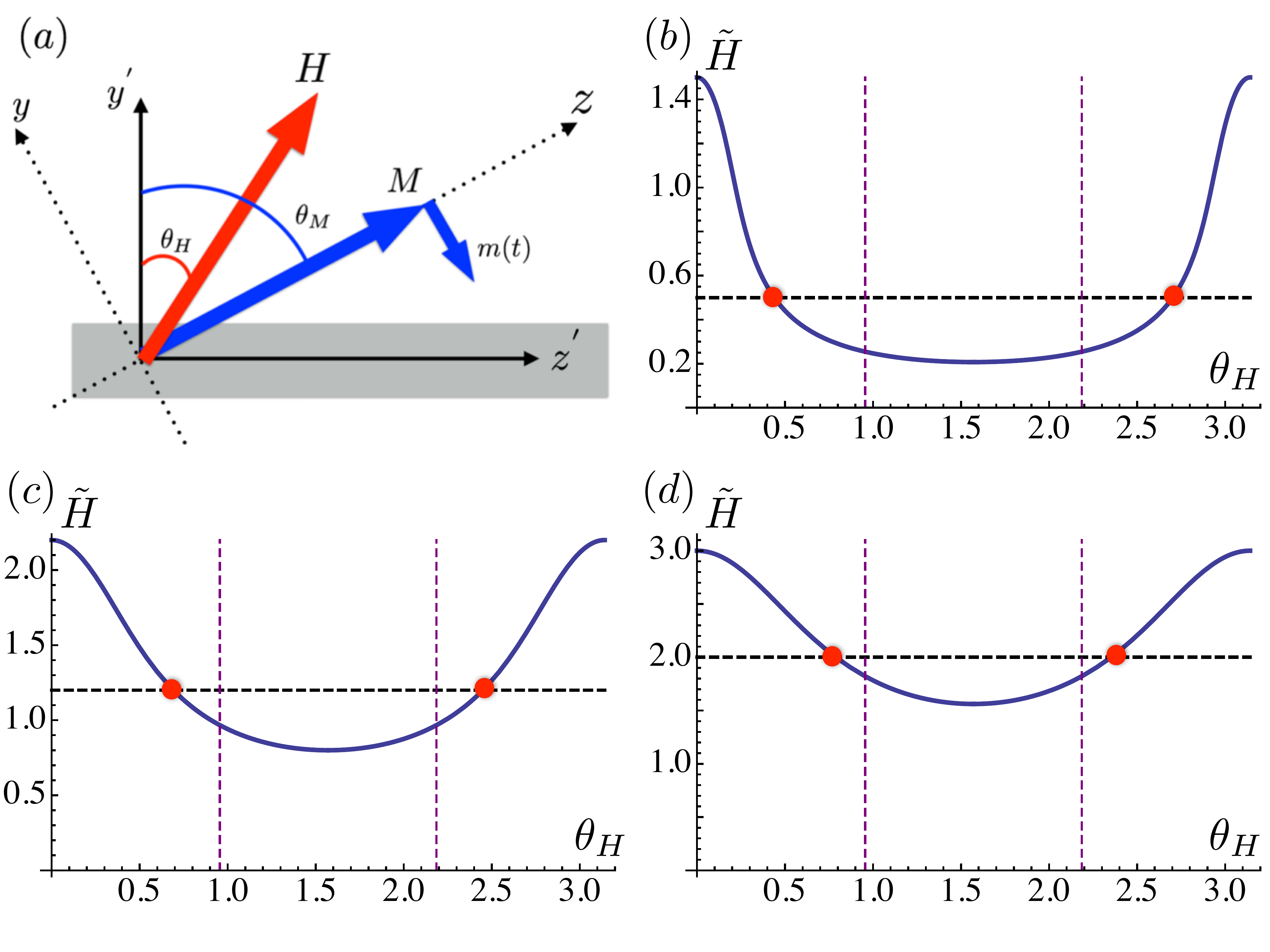}
\caption{(Color online) (a) The geometry of FMR; For a fixed dimensionless frequency,
$\tilde{\omega}=\omega_{\s \textit{FMR}}/4\pi M_s$, the dimensionless magnitude, ${\tilde H}=H/4\pi M_s$,  and orientation,
$\theta_H$, of dc magnetic field are related via Eq. (\ref{2}). This dependencies are shown for the values of $\tilde{\omega}/\gamma$:
(b) $0.5$, (c) $1.2$, and (d) 2. Red dots indicate the values of $H$, for which the
condition $\gamma H= \omega_{\s \textit{FMR}}$ is satisfied.        }
\label{Fig3}
\end{figure}

To generalize Eq. (\ref{current}) to a finite magnetic field in the normal layer, one must take into account the modification of the spin eigenstates, as well as the Zeeman splitting in energies of the latter. The spin structure of the spin-split levels depends on the orientation of ${\bm H}$ as follows
\begin{align}
|\chi_{\s {\bf H}+}\rangle=\cos\Bigl(\frac{\theta_H-\theta_M}{2}\Bigr)|\chi_{\s{\bf M}+}\rangle
+i\sin\Bigl(\frac{\theta_H-\theta_M}{2}\Bigr)|\chi_{\s {\bf M}-}\rangle,\\
|\chi_{\s {\bf H}-}\rangle=\cos\Bigl(\frac{\theta_H-\theta_M}{2}\Bigr)|\chi_{\s {\bf M}-}\rangle
+i\sin\Bigl(\frac{\theta_H-\theta_M}{2}\Bigr)|\chi_{\s {\bf M}+}\rangle.
\end{align}
Here the quantization axes for $|\chi_{\s{\bf M}\pm}\rangle$ and $|\chi_{\s{\bf H}\pm}\rangle$ spinors are chosen along the static part of the magnetization, and the external magnetic field, respectively, see Fig. \ref{Fig3}(a). The states $|\chi_{\s{\bf M}\pm}\rangle$ at sites $i$ and $j$ are split by $\Delta_z$.

All four transitions between states with $|\chi_{\s{\bf M}\pm}\rangle$ spin wave functions, Fig. \ref{Fig2}(b), are allowed for a general orientation of the magnetic field. For spin-conserving transitions ($+\rightarrow +$ and $-\rightarrow -$), the frequency dependence of $I^{(s)}$ remains $\omega r_{\omega}^2$, i.e. the same as in Eq.~(\ref{ZeroField}). Orientation of ${\bm H}$ enters into the prefactor: The product $m_x m_y$ should be replaced with $\frac{1}{4}\sin^2(\theta_H-\theta_M)m_x^2$ for both transitions.

While the spin-conserving transitions do affect the spin current density distribution in the sample, they are non-resonant, and it is the spin-flipping ones ($+\leftrightarrow -$) that are responsible for the spin current generation at the interface. In other words, no spin current is possible in a stationary state without the latter processes. Therefore, in what follows we concentrate on the frequency and magnetic field dependence of the corresponding rates.

As far as $+\rightarrow -$ and
$- \rightarrow +$ transitions are concerned, only the $+\rightarrow -$ with absorption of a magnon, and $-\rightarrow +$ with emission of a magnon  become important in the vicinity of the resonance $\hbar\omega=\Delta_z$. The other two transitions are non-resonant, and therefore disregarded here. For the $+\leftrightarrow -$ transitions, the prefactor $\omega$ in the spin current remains intact, since it comes from the difference in the populations of levels involved. However, despite the upper and lower Zeeman levels being separated in energy, the overlap of the spatial wave functions is determined by $\varepsilon_i$, $\varepsilon_j$ in {\em zero} magnetic field. Thus, the $+\rightarrow -$  transitions take place between pairs with $(\varepsilon_j-\varepsilon_i)\sim |\hbar\omega-\Delta_z|$.
These pairs have the ``shoulder"
\begin{equation}
\label{resonant}
  r_{\hbar\omega-\Delta_z}= a\ln\frac{2{t}_0}{|\hbar\omega-\Delta_z|}.
\end{equation}
Logarithmic divergence of Eq. (\ref{resonant}), which is cut off at $|\hbar\omega-\Delta_z|\sim \hbar/\tau$, ensures the resonant character of spin-flipping transitions that we took into account.

In addition to the replacement of $r_{\omega}$ by $r_{\omega-\Delta_z}$ in the spin current, the prefactor $m_xm_y$ should be modified as  $m_x m_y \rightarrow {\bf\mathrm{G}}(m_x,m_y)$, where the function ${\bf\mathrm{G}}$ is defined as
%\begin{align}
%\label{m}
%{\bf\mathrm{G}}(m_x,m_y)= & \frac{1}{4}(m_x^2+m_y^2)\bigl(\cos(\theta_H-\theta_M)+3\bigr)\\ &+2m_xm_y\cos(\theta_H-\theta_M)\nonumber,
%\end{align}
\begin{eqnarray}
\label{m}
{\bf\mathrm{G}}(m_x,m_y) = & \frac{1}{4}(m_x+m_y \cos(\theta_H-\theta_M))^2,
\end{eqnarray}
so that the absorption, and thus the FMR damping, do not have the usual form $\propto m_xm_y$.

\begin{figure}
\includegraphics[width=85mm]{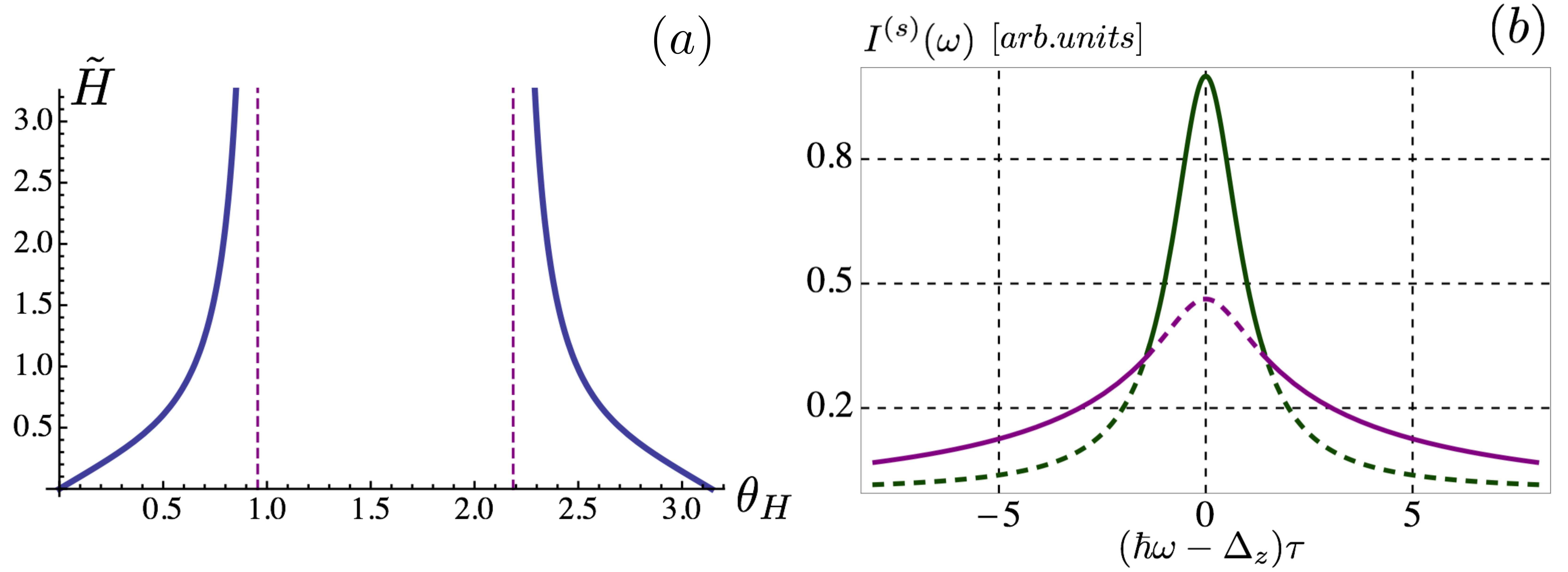}
\caption{(Color online) (a) the resonant condition $\gamma H=\omega_{\s \textit{FMR}}$
is satisfied along the solid lines on the $\left(\frac{H}{4\pi M_s}\right)-\theta_H$ plane. The cutoff values of
$\theta_H$ are $\cos^{-1}\left(\frac{1}{\sqrt 3}\right) \approx 55^{\circ}$ and $\pi-\cos^{-1}\left(\frac{1}{\sqrt 3}\right)$.
%(b) and (c) the dimensionless square
%of the matrix element $|M^2|$, Eq. ({\bf 6}), versus the orientation, $\theta_H$, of the ``resonant" magnetic field are plotted for the processes $\downarrow \rightarrow \uparrow$
%and $\uparrow \rightarrow \downarrow$, respectively.
(b) the behavior of the spin current calculated from Eq. (\ref{final}) for $ga^3\hbar/\tau=4\cdot 10^{-3}$ and ${t}_0\tau/\hbar=15$.}
\label{Fig4}
\end{figure}

The most spectacular manifestation of the resonance $\hbar\omega=\Delta_z$ is that the {\em intrasite} transitions become possible, as illustrated in Fig. \ref{Fig2}(b). For these transitions the overlap of the spatial parts of the on-site wave functions is equal to $1$, and the magnetic-field dependence of absorption is a pure Lorentzian. Orientation-dependent prefactor, which is the matrix element of $\delta \hat {\cal H}_i$ between the spinors $|\chi_{\s {\bf H}+}\rangle$ and $|\chi_{\s {\bf H}-}\rangle$ is the same as in Eq.~(\ref{m}).
Summarizing, we present the expression for spin current close to the resonance $\hbar\omega=\Delta_z$ in the form
\begin{multline}
\label{final}
I^{(s)}(\omega)=2{\bf\mathrm{G}}(m_x,m_y)J^2\omega\frac{\partial f}{\partial \varepsilon}\\ \times
\Bigl[\frac{\frac{\hbar}{\tau}}{(\Delta_z-\hbar\omega)^2+\left(\frac{\hbar}{\tau}\right)^2}
+\pi^2 ga^3\ln^2\frac{2{t}_0}{|\hbar\omega-\Delta_z|}\Bigr],
\end{multline}
where the first term comes from intrasite and the second term from intersite transitions.
Directly at the resonance, the first term dominates. This is ensured by the condition
$ga^3\hbar/\tau\ll 1$. Since the combination $1/ga^3$ is the minimal energy spacing between two
sites in the insulator located within $\sim a$ from each other,
%nearest neighbors in the insulator,
the above condition implies that this spacing is much bigger
than the phonon broadening of individual levels, which is the definition of the Anderson
insulator. As the deviation from the resonance increases, the behavior of $I^{(s)}(\omega)$ is
dominated by the second term. Neglecting the logarithm, the crossover takes place at
$|\Delta_z-\hbar\omega|\tau/\hbar \gtrsim \left(\tau/\hbar ga^3\right)^{1/2}\gg 1$. The behavior of spin current
near the resonance is shown in Fig.~\ref{Fig4}(b), where the logarithm was cut off at $|\hbar\omega-\Delta_z|=t_0/15$.

\section{Resonant orientations of external field}

Equation~(\ref{final}) is our main result. To make connection to the experimental papers
Refs.~[\onlinecite{Polaron1,Polaron2,Polaron3,Polaron4}], below we calculate the magnitude and orientation of the dc field where the anomalous behavior of ISHE voltage takes place. Such behavior takes place when two conditions are met: The Zeeman splitting of the localized states is equal to $\hbar \omega$, and $\omega=\omega_{\s \textit{FMR}}$.

We specify the orientation of ${\bm H}$ and magnetization, ${\bm M}$,
using the notations common in the literature, see e.g. Refs.~[\onlinecite{Pt6,silicon1,Polaron2}], and Fig.~\ref{Fig3}. We will also introduce dimensionless variables
${\tilde H}$, ${\tilde M}$ and ${\tilde \omega}$, which stand for
 $H$, $M$ and $\omega_{\s \textit{FMR}}$
in the units of $4\pi M_s$, where $M_s$ is the saturation magnetization. Then the angle $\theta_M$, corresponding the equilibrium orientation of ${\bm M}$, is found from the condition that ${\bm M}$
is parallel to the effective magnetic field, with the demagnetizing term taken into account\cite{Pt6}

\begin{equation}
\label{1}
2\tilde{H}\sin(\theta_H-\theta_M)+\sin2\theta_M=0,
\end{equation}
while the expression for the resonant frequency, ${\tilde\omega}$, reads\cite{Kittel}
\begin{multline}
\label{2}
\Big(\frac{\tilde{\omega}}{\gamma}\Big)^2=\Big[\tilde{H}\cos(\theta_H-\theta_M)-\cos2\theta_M\Big]
\\\times \Big[\tilde{H}\cos(\theta_H-\theta_M)-\cos^2\theta_M\Big].
\end{multline}
From these two equations we exclude $\theta_M$ and plot the dimensionless field
$\tilde{H}$ versus $\theta_H$, for a {\em given} FMR frequency $\tilde{\omega}$. Examples
of these curves are shown in Fig. \ref{Fig3}. Resonant orientation is obtained by crossing
a curve $\tilde{H}(\theta_H)$ by the line $\tilde{\omega}=\gamma\tilde{H}$. Two intersections determine
the orientations for which $\omega_{\s \textit{FMR}}$ is equal to the Zeeman splitting of the localized states.
Upon changing $\omega_{\s \textit{FMR}}$, we get two lines of resonances, Fig. \ref{Fig4}(a). They occupy two domains:
$0< \theta_H <\cos^{-1}\frac{1}{\sqrt{3}}$ and $(\pi-\cos^{-1}\frac{1}{\sqrt{3}})< \theta_H<\pi$.
At the boundaries of the domains $\tilde{H}$ goes to infinity. Then it follows from Eqs. (\ref{1}) and
(\ref{2}) that at these boundaries $\sin(\theta_M-\theta_H)=0$, and $\theta_H$  satisfies   the  equation $\cos(2\theta_H)+\cos^2\theta_H=0$, yielding
$\theta_H =\cos^{-1}\left(\frac{1}{\sqrt 3}\right)\approx 55^{\circ}$.

In Refs.~[\onlinecite{Polaron1,Polaron2}] on pumping into organics
the microwave frequency driving the resonance was $9.45$~Ghz,
while the values of $4\pi M_s$ were very different, namely, $4\pi M_s=0.175$ T~in Ref.~[\onlinecite{Polaron1}]
and $4\pi M_s=0.805$~T in Ref.~[\onlinecite{Polaron2}]. Then from Fig.~\ref{Fig4}(a) we find that the resonant angle $\theta_H$ should be close to $45^{\circ}$ for Ref.~[\onlinecite{Polaron1}] and $23^{\circ}$
for Ref.~[\onlinecite{Polaron2}].

\begin{figure}
\includegraphics[width=45mm]{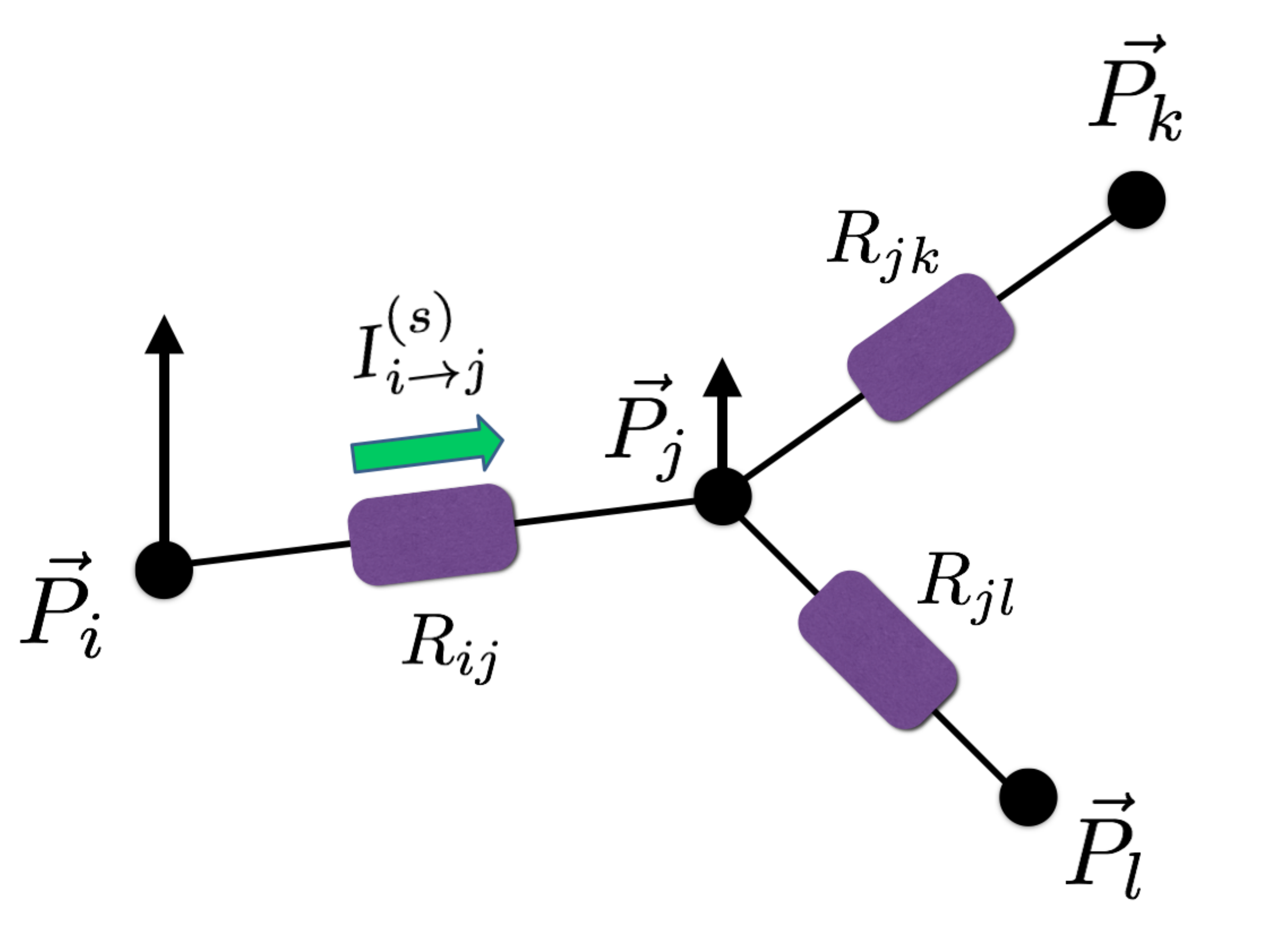}
\caption{(Color online) ``Spin-resistor" network.  Polarizations ${\bm P}_i$ and ${\bm P}_j$ on the sites $i$, $j$
determine the spin current between these sites. The coefficients, $R_{ij}$, are proportional
to the electric hopping resistances.}
\label{Fig5}
\end{figure}

\section{Spin-resistor network}
\label{SCN}
After the spin polarization is generated at the boundary, it should spread into the bulk of
the insulator  to avoid the
backflow.\cite{flowback} In a metal, where ${\bm P}$ is a continuous function of coordinates, this spreading is by spin diffusion accompanied by the Larmor precession.
In a hopping insulator  ${\bm P}$ takes discrete values, ${\bm P}_i$, which are the polarizations on the sites, $i$. The Larmor precession is accounted for by the on-site Zeeman splitting, $\Delta_z$, of the levels, see Fig. \ref{Fig2}. The frequencies of electron hops between two sites, $i$ and $j$, depend strongly on their energies, $\varepsilon_i$, $\varepsilon_j$, and their spatial separation, $r_{ij}$. Then the issue of spreading of the spin polarization reduces to the question: what is the spin current ${\bm I}^{(s)}_{i\rightarrow j}$ between the sites with polarizations  ${\bm P}_i$ and ${\bm P}_j$, provided that, on average, there is no charge current between these sites?

If bias were applied between the two sites, then the average charge current, proportional to this bias, could be found
by ascribing an effective resistance, $R_{ij}$, to the pair of sites\cite{HalperinLanger,book}. It is easy to see that the same $R_{ij}$ determines the proportionality coefficient between ${\bm I}^{(s)}_{i\rightarrow j}$ and ${\bm P}_i-{\bm P}_j$, namely
\begin{equation}
\label{network}
{\bm I}^{(s)}_{i\rightarrow j}=2\frac{{\bm P}_i-{\bm P}_j}{R_{ij}\frac{\partial f}{\partial \varepsilon}}.
\end{equation}
In Eq. (\ref{network}) we have assumed that the difference  ($\varepsilon_i - \varepsilon_j$) is much smaller than the temperature, so that $\frac{\partial f}{\partial \varepsilon}$ is
the same for both sites. Equation (\ref{network}) follows from the fact that the on-site chemical potentials of the local majority and minority electrons are shifted by  $\mp|{\bm P}_i|/\frac{\partial f}{\partial \varepsilon}$, respectively. The spinors that correspond to these local spin eigenstates are defined by
the directions of ${{\bm P}_i}$, ${{\bm P}_j}$. Importantly, the fact that the chemical potential splitting is symmetric around the chemical potential of the unpolarized system ensures the absence of the charge current, i.e. the net current flow $i\rightarrow j$ is compensated by the net current flow $j\rightarrow i$.
With different spin polarizations of the sites, the compensation of the charge flows leads to the
imbalance of the spin flows, and thus to Eq.~(\ref{network}).
%If both sites are spin-polarized, the corresponding flows
%are also spin-polarized. This leads to  Eq.~(\ref{network}).
Note that Eq.~(\ref{network}) remains valid in  external magnetic field, which enters only via the magnitudes of polarizations. Overall, Eq.~(\ref{network}) suggests that polarization built up at the F-N boundary spreads along the current-carrying resistor network, as illustrated in Fig. \ref{Fig5}.

%\section{Absorption of magnons at F-N boundary}

%\vspace{5mm}

\section{Concluding remarks}

\noindent (i) Our result Eq. (\ref{final}) applies when the phonon-induces broadening of the levels is smaller than $\omega$. In the opposite case, $\omega\tau \ll 1$, the mechanism of absorption is the Pollak-Geballe relaxation mechanism, Ref.~[\onlinecite{PollakGeballe}]; no sharp dependence of pumping near the resonance
is expected in this regime. Unlike pumping into metals, the pumping rate Eq.~(\ref{final}) is not simply proportional to ${\bm m}\times \frac{d{\bm m}}{dt}$. The prefactor in Eq. (\ref{m}) depends on the relative
orientation of ${\bm m}$ and the external magnetic field, ${\bm H}$.

\noindent (ii)
Suppose that we are at resonance $\hbar\omega=\Delta_z$. The microwave field acts both by driving the FMR
 but also directly, by causing transitions between the Zeeman levels. If the amplitude of the field in frequency units (Rabi frequency) exceeds the inverse spin relaxation time, these transitions will be saturated in the bulk. Then the pumping becomes inefficient.

\noindent (iii) In  conventional theory of hopping transport the applied voltage drops
not on all the resistors constituting the network, but on the highest, critical, resistors
representing the ``hardest" hops.\cite{book} The spin relaxation rate will be dominated by
hyperfine or spin-orbit environment\cite{Flatte,Roundy} of this hop.

\noindent (iv) We did not consider effects of electron-electron interaction, and did not describe in detail how finite resistance of the spin-current network, Section~\ref{SCN}, affects the measured value of the spin current. In brief, Coulomb correlations enhance the absorption of magnons by increasing the number of singly occupied pairs\cite{Efros}, while the measured spin current is given by Eq.~(\ref{final}) only in the limit of vanishingly small bulk resistance. These considerations are, however, completely standard, and do not change the qualitative picture of spin current generation by resonant magnon absorption in hopping insulators, developed in this paper.

%\section{ Acknowledgements}
\noindent{\em Acknowledgements.}
We are grateful to  C. Boehme  and Z. V. Vardeny for piquing our interest in the subject. The work was supported by NSF grants MRSEC DMR-1121252 (ZY and MER), and DMR-1409089 (DAP).


\begin{thebibliography}{99}
\bibitem{ActivationEnergy}
For example, the bulk resistance, $\rho(T)$, in Ref.~[\onlinecite{Polaron4}] increased by two orders
of magnitude between the room temperature and $20$~K following the Mott's law $\ln \rho \propto T^{-1/4}$.
The localization length  inferred by the authors from the transport measurements is
quite small: $a \approx 11$ nm.
\end{thebibliography}
\end{document}